\documentclass[aps,prd,showpacs,floatfix,nofootinbib,superscriptaddress,twocolumn,10pt]{revtex4}
\usepackage{graphicx}
\usepackage{bm}
\usepackage{subfigure}
\usepackage{amssymb}
\usepackage{color,soul}
\usepackage{graphicx}
\usepackage{multirow}
\graphicspath{{figure/}}
\DeclareGraphicsExtensions{.pdf,.png,.jpg}
\setulcolor{red}

\newcommand{\beq}{\begin{equation}}
\newcommand{\eeq}{\end{equation}}
\newcommand{\beqa}{\begin{eqnarray}}
\newcommand{\eeqa}{\end{eqnarray}}

\begin{document}

\title{Plasma dark matter and electronic recoil events in XENON1T}

\author{Lei Zu}
\affiliation{Key Laboratory of Dark Matter and Space Astronomy, Purple Mountain Observatory, Chinese Academy of Sciences, Nanjing 210023, China}
\affiliation{School of Astronomy and Space Science, University of Science and Technology of China, Hefei, Anhui 230026, China}

\author{R. Foot \footnote {Corresponding author: rfoot@unimelb.edu.au}}
\affiliation{School of Physics, University of Melbourne, Victoria 3010 Australia}
\affiliation{School of Physics, University of Sydney, NSW 2006, Australia}

\author{Yi-Zhong Fan \footnote{Corresponding author: yzfan@pmo.ac.cn}}
\affiliation{Key Laboratory of Dark Matter and Space Astronomy, Purple Mountain Observatory, Chinese Academy of Sciences, Nanjing 210023, China}
\affiliation{School of Astronomy and Space Science, University of Science and Technology of China, Hefei, Anhui 230026, China}

\author{Lei Feng \footnote{Corresponding author: fenglei@pmo.ac.cn}} 
\affiliation{Key Laboratory of Dark Matter and Space Astronomy, Purple Mountain Observatory, Chinese Academy of Sciences, Nanjing 210023, China}
\affiliation{Joint Center for Particle, Nuclear Physics and Cosmology,  Nanjing University -- Purple Mountain Observatory,  Nanjing  210093, China}

\begin{abstract}
Dark matter might be in the form of a dark plasma in the Milky Way halo.
Specifically, we consider here a hidden sector consisting of a light `dark electron' and a much heavier `dark proton',
each charged under an unbroken $U(1)'$ gauge symmetry.
These self-interacting dark sector particles can also interact with ordinary matter 
via the kinetic mixing interaction, and lead to a signal in dark matter direct detection experiments. 
Indeed, keV electron recoils can arise quite naturally in such models from dark electron scattering off loosely bound atomic electrons.
Here we examine the recently reported XENON1T excess in the context of such a plasma dark matter model. 
We find that the observed excess can be explained if kinetic
mixing is in the approximate range: $10^{-12} \lesssim \epsilon \lesssim 10^{-10}$.
The allowed parameter space is consistent with astrophysical and cosmological constraints 
and consistent also with other direct detection experiments.

\end{abstract}

\maketitle

\section{Introduction}

Over the last few decades, dark matter experiments have set stringent limits on dark matter interactions with both nuclear and 
electronic components, e.g. \cite{Akerib,Cui,aprile,Agnes}.
Recently the XENON1T experiment has reported an excess of low energy electron recoil events in an S1 analysis near their $\sim 2$ keV 
threshold \cite{xenon1t}. Although the significance of this excess is only around $3.5 \sigma$, these results have sparked much interest, 
including interpretations involving axions e.g. \cite{Tak,Sun,Cai,axion,Gao}, non-standard neutrino interactions e.g. \cite{Khan,Bally,french,val}, 
WIMP dark matter e.g. \cite{strum,Dey,Fornal} and more exotic scenarios e.g. \cite{bell,yas}. These interpretations are not without challenges. 
For example, the solar axion and neutrino magnetic moment could provide a good fit to the spectrum but the parameters required appear to be in
tension with stellar cooling constraints, while many of the proposed WIMP explanations require the velocity of the dark matter particles to exceed 
the escape velocity of the Milky Way.

Actually, electron recoils in the keV range can occur quite naturally in a class of hidden sector dark matter models which feature a 
dark plasma consisting of light and heavy components coupled together via a massless dark 
photon that is also kinetically mixed with the ordinary photon \cite{plasmadm,footyyy}. 
The underlying reason is straightforward:
Energy equipartition implies that the light component velocities are much larger than would be expected from
single component virialisation. In fact the light component can have typical velocities which are much larger than the galactic
escape velocity. These particles are bound to the galaxy halo not by gravity but by the dark electromagnetic force.

A theoretically constrained plasma dark matter model of this kind is mirror dark matter (for a review see \cite{freview}), 
which features a hidden sector exactly isomorphic to the Standard model \cite{flv}.
In that case the light component is the mirror electron, whose mass is constrained to be exactly the same as the electron due to an unbroken $Z_2$ symmetry. 

Although conceptually simple, mirror dark matter
is quite nontrivial with regards to direct detection experiments due to the influence
of captured dark matter within the Earth. 
The Earth-bound mirror helium distribution extends to the Earth's surface and 
is expected to drastically modify the distribution locally due to 
shielding effects \cite{plasmadm,shield1,shield2}. 
An interpretation of the XENON1T excess as mirror electron scattering
off loosely bound atomic electrons has been suggested in \cite{Zu}. In that work, it was assumed that 
only the high velocity tail of the halo mirror electron distribution was able to reach the detector due to shielding.
It was shown there that the XENON1T excess can be simply explained
in that model, consistently with other experimental constraints including the Darkside-50 constraint \cite{Agnes}.

In this paper, we examine the more general two component model consisting of a dark proton and dark
electron whose masses are theoretically unconstrained. 
In this case, captured dark matter within the Earth can be quite compact in the limit where the dark proton mass is large ($> 20$ GeV). In fact, 
we will consider the idealized limiting case where the influence of this captured dark matter on the halo dark electron distribution at the detector is assumed 
to be unimportant and the 
distribution is Maxwellian. 
While such  assumptions are unlikely to be strictly valid, this analysis should serve to illustrate 
the plasma dark matter interpretation of the XENON1T excess, and hopefully provide a useful `ball park' estimate for the parameter space of interest.

\section{The model}

Dark matter might arise from a hidden sector with at least two massive components coupled together via an unbroken $U(1)'$ gauge symmetry.
In that case, dark matter can form a plasma in the Milky Way halo \cite{vagnozzi}. Consider a two component model, consisting of a light dark electron 
and a heavy dark proton, assumed for now to have equal and opposite $U(1)'$ charge. 

The Lagrangian describing the standard model extended with such a hidden sector takes the form:
\begin{equation}
{\cal L} = {\cal L}_{SM}\ + \ {\cal L}_{dark}\ + \ {\cal L}_{mix}
\label{lagrangian}
\end{equation}
where
\begin{eqnarray}
{\cal L}_{dark} &=& -\frac{1}{4} F^{\prime \mu \nu} F^{\prime}_{\mu \nu} + \bar p_d (iD_\mu \gamma^\mu - m_{p_d})p_d \nonumber \\
      & + &
\bar e_d (iD_\mu \gamma^\mu - m_{e_d})e_d .
\end{eqnarray}
Here $D_\mu = \partial_\mu + ig'Q'A'_\mu$ is the dark $U(1)'$ covariant derivative.
That is, ${\cal L}_{dark}$ describes the dark electron and dark proton interactions with a massless dark photon, quite analogous to the electromagnetic
interactions of ordinary electrons and protons. There is a dark fine structure constant $\alpha_d = (g'Q'(e_d))^2/4\pi$.

The ${\cal L}_{mix}$ part describes the non-gravitational interactions coupling the dark sector particles to ordinary matter. Restricting to
renormalizabile interactions, there is just one term,  the kinetic mixing interaction \cite{foothe,holdom}:
\begin{equation}
{\cal L}_{mix}=\frac{\epsilon^{\prime}}{2} F^{\mu \nu}F^{\prime}_{\mu \nu}
\label{l_mix}
\end{equation}
where $F^{\mu\nu}$ [$F^{\prime}_{\mu\nu}$] is the ordinary [dark] $U(1)_Y$ [$U(1)'$] gauge boson field strength tensor. 
The effect of this kinetic mixing is to induce tiny ordinary electric charges of $\pm \epsilon e$ for the dark proton and dark electron.
For illustration we assumed above that the $e_d, p_d$ $U(1)'$ charges are equal and opposite, the most general case has a charge
ratio parameter, $Z' \equiv Q'(p_d)/Q'(e_d)$. The physics of the most general two component model is fully described by five parameters:
$m_{e_d}, m_{p_d}, \alpha_d, \epsilon, Z'$ \cite{vagnozzi}.

In this class of dark sector models, the kinetic mixing interaction can play an important role in cosmology and galaxy structure. 
Indeed, it has been argued that successful small scale structure implicates kinetic mixing strength of order 
$\epsilon \sim 10^{-10}$ \cite{vagnozzi2}. Such small values are consistent with other constraints, including early Universe
cosmology, e.g. BBN constraints \cite{vagnozzi,other}.

The kinetic mixing interaction is of course important also for direct detection experiments. 
It facilitates Rutherford-type scattering of dark electrons off loosely bound atomic electrons. For parameter space with 
sufficiently high halo temperature,  electron recoils in the keV range arise, and have been investigated quite extensively with regards 
to direct detection experiments over the last few years \cite{plasmadm,shield1,shield2}.
It has been suggested that it might even be possible to 
explain the DAMA annual modulation signal \cite{dama} (and reference there-in) in this
manner, since the constraints on electron recoils provided by other experiments are
generally much weaker than those on nuclear recoils. That DAMA explanation predicted a low energy keV electron recoil excess \cite{shield1,shield2}
somewhat larger than what has been observed in the XENON1T experiment. 
However, at the present time it is unclear if that DAMA explanation is robustly excluded due to potential
uncertainties in modelling the falling detection efficiency in XENON1T, as well as resolution and energy-scale uncertainties in both XENON1T and DAMA experiments.
There are also significant constraints from the LUX experiment, albeit at a different geographical location \cite{luxc}.

There is some restriction on parameter space from galactic considerations \cite{plasmadm,vagnozzi}, which we summarize below for the case where the
dark proton and dark electron have equal and opposite $U(1)'$ charge (i.e. $Z' = 1$).
The parameter region where the dark matter takes the form of a plasma (i.e. is highly ionized) in the  Milky Way dark halo is 
\begin{eqnarray}
\left( \frac{\alpha_d}{10^{-2}}\right)^2\left( \frac{\mu_d}{{\rm MeV}} \right)\left(\frac{{\rm GeV}}{m_{p_d}}\right) \lesssim 1
\end{eqnarray}
where $\mu_d \equiv m_e m_{e_d}/(m_e + m_{e_d})$ is the reduced mass.
Since the dark halo is dissipative it can be potentially unstable (i.e. collapse onto a disk). In the absence of heating
a lower limit on $m_{p_d}$ can be estimated by requiring the cooling time-scale be longer than the Hubble time. For the Milky Way this 
condition yields the estimate:
\begin{eqnarray}
m_{p_d} \gtrsim 20 \left( \frac{{\rm MeV}}{m_{e_d}}\right) \left(\frac{\alpha_d}{10^{-2}}\right)^2 \ {\rm GeV}.
\label{cov}
\end{eqnarray}
In the presence of a halo heating mechanism, this bound can be relaxed. If $\epsilon \sim 10^{-10}$ then substantial heating of the halo can
be provided by type II  supernovae \cite{fv}. That scenario has been studied in detail in the mirror dark matter case
(as the condition, Eq.(\ref{cov}), is violated in that model) \cite{f2000} (and references there-in).

An upper limit on $m_{p_d}$ can be estimated by requiring the energy transfer between the light dark electrons and heavy dark protons
be efficient enough so that energy equipartian is approximately valid. An estimate assuming that this energy transfer is 
dominated by two-body collisional processes gives 
\begin{eqnarray}
m_{p_d} \lesssim 200 \left( \frac{m_{e_d}}{{\rm MeV}}\right)^{1/7} \left( \frac{ \alpha_d}{10^{-2}}\right)^{4/7} \ {\rm GeV} .
\end{eqnarray}
This limit should be viewed as conservative as collective plasma effects can also play a role but are more difficult to estimate.

If energy equipartian is indeed approximately valid then the halo dark electrons and dark protons have a common temperature, 
estimated to be \cite{plasmadm,vagnozzi} :
\begin{equation}
T \approx \frac{1}{2} \bar{m} v^{2}_{rot},
\label{temperature}
\end{equation}
where $\bar{m}= (m_{p_d} + m_{e_d})/2$ is the mean particle mass in the (assumed) fully ionized halo,
and $v_{rot}$ is the galactic rotational velocity ($\sim 220$ km/s for 
the Milky Way).\footnote{Units where $k_B = \hbar = c =1$ are assumed unless otherwise indicated.}
In the limit where the dark electrons are much lighter than the dark protons,
typical dark electron velocities in the Milky Way halo can be much larger than the galactic escape velocity (as calculated from gravity). 
The dark electrons do not escape the galaxy due to $U(1)'$ neutrality; the plasma is highly
conducting, and dark electric forces keep the plasma neutral over length scales larger than the Debye length.

This kind of dark matter can be captured in the Earth where it thermalizes with the ordinary matter and forms an extended distribution.  This captured dark
matter can potentially strongly influence the halo dark matter distribution arriving at a detector. This dark sphere of influence is caused
by the generation of dark electromagnetic fields in both the captured dark matter and halo plasma near the Earth \cite{plasmadm} and potentially
also to collisional shielding effects \cite{shield1}.
An estimate for the physical extent of the captured dark matter distribution within the Earth yields \cite{plasmadm,fv15}
\begin{eqnarray}
\frac{R_{DM}}{R_E} \approx \left( \frac{5 \ {\rm GeV}}{m_{p_d}/\zeta}\right)^{0.55} \left( \frac{\alpha_d}{10^{-2}}\right)^{0.06}
\label{888}
\end{eqnarray}
for $5\times 10^{-4} \lesssim \alpha_d \lesssim 5\times 10^{-2}$, $5\ {\rm GeV} \lesssim m_{p_d} \lesssim 300 \ {\rm GeV}$ and
$\zeta$ is in the range, $1 < \zeta < 2$, depending on the ionization state of the captured dark matter within the Earth.
In the limit where $R_{DM}/R_E \to 0$ the influence of captured dark matter on the halo dark matter distribution near the Earth's surface is
expected to become negligible. This idealized limiting case will be assumed in this work, which Eq.(\ref{888})
suggests should be useful for sufficiently large $m_{p_d}$, taken to be $m_{p_d} \gtrsim 20$ GeV.

\section{Dark electron and dark proton interactions with matter}

Dark electrons and dark protons interact with ordinary matter due to the kinetic mixing interaction, Eq.(\ref{l_mix}).
Taking the illustrative  case where the dark electron and dark proton have equal and opposite $U(1)'$ charge,
the kinetic mixing interaction endows a small effective electric charge for the dark electron and dark proton of $\pm \epsilon e$.
The kinematically important channels are dark electron - electron scattering and dark proton - nuclei scattering.

Coulomb scattering of a dark electron off an electron
is a spin-independent process. Approximating the target electron as free and at rest
relative to the incoming dark electron of velocity $v$, the cross section in the non-relativistic limit is:
\begin{equation}
\frac{d\sigma}{dE_R}=\frac{\lambda}{E_R^2|v|^2},
\label{crosssection}
\end{equation}
where
\begin{equation}
\lambda=\frac{2\pi\epsilon^2\alpha^2}{m_e}.
\label{lambda}
\end{equation}
Here $E_R$ is the recoil energy of the target electron and $\alpha$ is the fine structure constant. Considering bound target electrons as free 
could only be a useful approximation for loosely
bound atomic electrons, i.e. those with binding energy much less than $E_R$. 
For a xenon atom there are $g = 44$ such loosely bound electrons in the recoil energy region of interest.


The Earth is moving with respect to the dark matter distribution with some speed, $v_B \sim 220$ km/s. If we denote by $f_{e_d} (v;v_B)$
as the dark electron distribution in the Earth's reference frame, then the local differential rate is:
\begin{eqnarray}
\label{differential}
\frac{dR}{dE_R}&=&gN_{T}n_{e_d}\int \frac{d\sigma}{dE_R} f_{e_d}(v;v_B) |v|d^3v  \nonumber\\
&=& gN_{T}n_{e_d} \frac{\lambda}{E_R^2} \int_{|v|>v_{min}}^\infty \frac{f_{e_d}(v;v_B)}{|v|}d^3v,
\end{eqnarray}
where
\begin{equation}
v_{min}=\frac{\sqrt{m_e E_R/2}}{\mu_d}.
\label{v_min}
\end{equation}
Here $\mu_d \equiv m_em_{e_d}/(m_e+m_{e_d})$ is the reduced mass, $n_{e_d}=\rho_0/(m_{e_d}+m_{p_d})$ denotes the number density of $e_d$ with the 
local dark matter density $\rho_0=0.3~{\rm GeV/cm^{3}}$ and $N_{T}$ is the number of target Xe atoms per tonne. 

The dark electron distribution arriving at the detector is of course uncertain. In the following we make the simple assumption of a
Maxwellian distribution, boosted by the velocity $v_B$:
\begin{eqnarray}
f_{e_d}(v;v_B) = \left( \frac{1}{\pi v_0^2} \right)^{3/2}  exp\left( \frac{-|v-v_B|^2}{v_0^2}\right)
\end{eqnarray}
where $v_0^2 \equiv 2T/m_{e_d}$.  From Eq.(\ref{temperature}), $v_0=\sqrt{\bar{m}/m_{e_d}} v_{rot}$ for dark electrons.
The differential rate evaluates to \cite{plasmadm}:
\begin{eqnarray}
\label{drde_witherf}
\frac{dR}{dE_R} = \frac{g N_T n_{e_d} \lambda}{2E_R^2 |v_B|} \left[ {\rm erf}(x+y) -
{\rm erf}(x-y) \right] 
\end{eqnarray}
where $x=v_{min}/v_0$ and $y=|v_B|/v_0$.
In the limit $y \ll 1$, expected to be valid for electron recoils (but not for nuclear recoils), the rate reduces to:
\begin{equation}
\frac{dR}{dE_R}=gN_Tn_{e_d}\frac{\lambda}{E_R^2}\left(\frac{2e^{-x^2}}{\sqrt{\pi}v_0}\right).
\label{drde}
\end{equation}
In the parameter region where $x \ll 1$, the rate of keV electron recoils evaluates to: 
\begin{eqnarray}
\frac{dR}{dE_R} &\approx &
8 \left( \frac{\epsilon}{10^{-12}} \right)^2 \left( \frac{m_{e_d}}{\rm MeV}\right)^{1/2}\left( \frac{20\ {\rm GeV}}{m_{p_d}}\right)^{3/2}
\left( \frac{ {\rm keV}}{E_R}\right)^2 \nonumber \\
&  &  \ \ \ \ \ \ \ \ \ \ \ \ \ \ \ \ \  \ \ \ {\rm tonne}^{-1} \ {\rm year}^{-1} \ {\rm keV}^{-1} .
\label{num7}
\end{eqnarray}

So far we have discussed the interactions of dark electron scattering of bound atomic electrons. Nuclear recoils can also be (potentially)
observable in direct detection experiments.
For dark proton scattering off nuclei (assuming a xenon target for definiteness) the cross section has a similar form to Eq.(\ref{crosssection}),
but with $\lambda$ replaced by:
\begin{equation}
\lambda_c=\frac{2\pi\epsilon^2Z^2\alpha^2}{m_{A}}F_A^2(qr_A)
\label{lambda_nuclear}
\end{equation}
where Z=54 and A=131 are the atomic and mass number for xenon, $m_{Xe}$ represents the mass of the xenon atom and $F_A (qr_A)$ 
is the form factor which takes into account the finite size
of the xenon nucleus. Here $q=(2m_AE_R)^{1/2}$ is the momentum transfer and $r_A$ is the effective nuclear radius. 
The form factor that we adopt is the analytic expression given by Helm \cite{helm,lewin}:
\begin{equation}
F_A(qr_A)=3\frac{j_1 (qr_A)}{qr_A}e^{-(qs)^2/2},
\label{form_factor}
\end{equation}
with $r_A=1.14A^{1/3} $ fm, s=0.9 fm and $j_1$ is the spherical Bessel function of index 1.
For nuclear recoils, the rate is then given by:
\begin{eqnarray}
\label{drde_nuclear}
\frac{dR}{dE_R} = \frac{N_T n_{p_d} \lambda_c}{2E_R^2 |v_B|} \left[ {\rm erf}(x+y) -
{\rm erf}(x-y) \right].
\end{eqnarray}
where $x=v_{min}/v_0$ and $y=|v_B|/v_0$.
Note that for dark protons,  $v_{min}$ is of the form Eq.(\ref{v_min}) but with $m_e \to m_A$, $m_{e_d} \to m_{p_d}$,
while $v_0 = \sqrt{\bar{m}/m_{p_d}}v_{rot} \approx v_{rot}/\sqrt{2}$, and hence
$y \approx \sqrt{2}$.

\section{results}

In order to compare with the experimental results, the detector energy resolution and detection efficiency needs
to be modelled. For the energy resolution, we convolved Eq.(\ref{drde}) with a Gaussian distribution and incorporated a detector
efficiency function, $\gamma(E)$:
\begin{equation}
\frac{dR}{dE_m}=\frac{1}{\sigma\sqrt{2\pi}}\int\frac{dR}{dE_R}e^{-(E_R-E_m)^2/2\sigma^2}\times \gamma(E_R) dE_R,
\label{final}
\end{equation}
where $E_m$ denotes the measured recoil energy in the experiment and $\sigma$ is the detector averaged resolution. 
At low electron recoil energies, the  S1 resulution is poorly constrained, and we assumed $\sigma = 0.5\ {\rm keV}$. 
The estimate of $\gamma(E)$ was obtained from Ref. \cite{xenon1t}.



\begin{figure}[htbp]

\subfigure[\ $m_{e_d}=1~{\rm MeV}, \, m_{p_d}=20~{\rm GeV}, \, \epsilon=6.8\times 10^{-12}$]{\includegraphics[width=0.90\columnwidth]{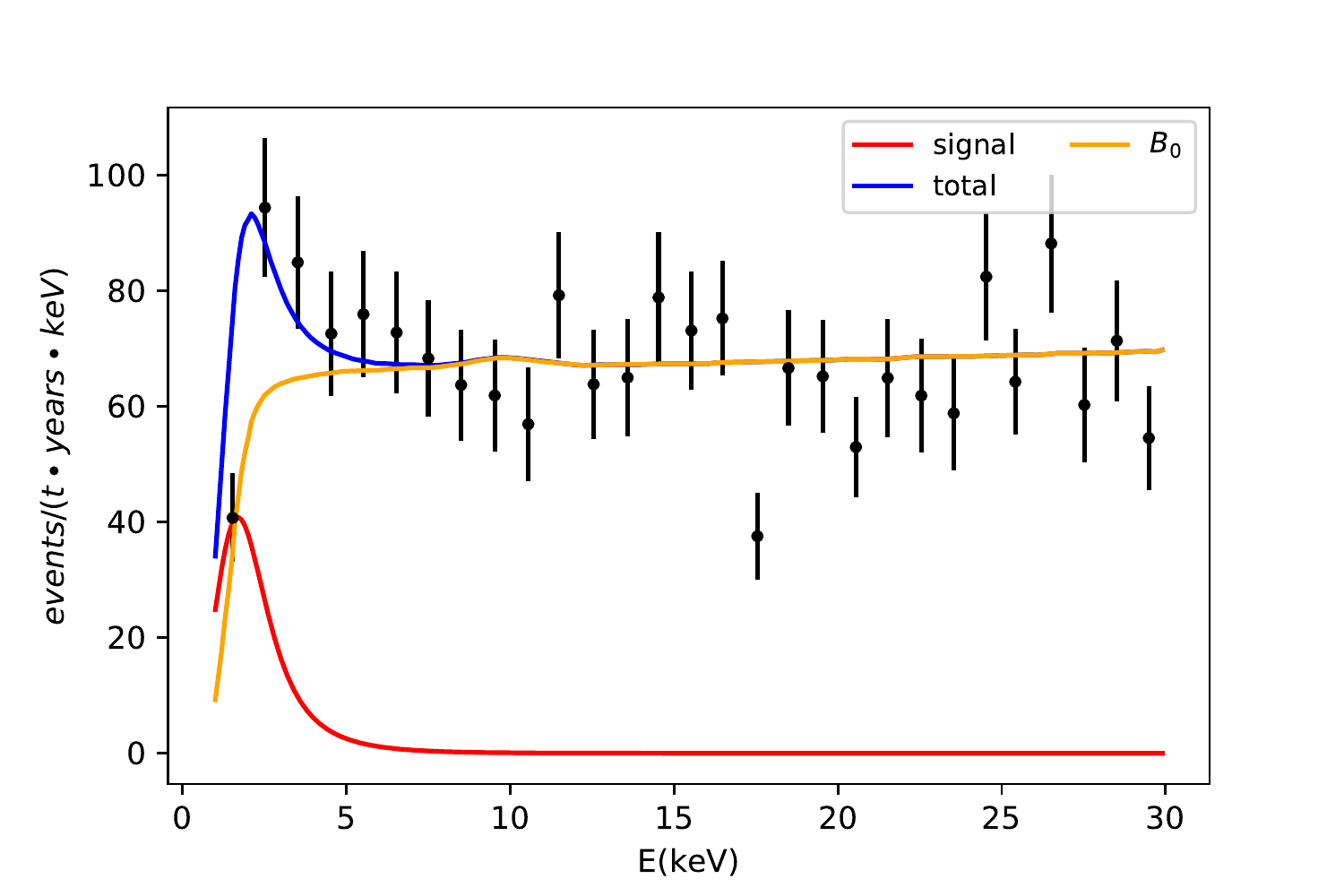}}

\subfigure[\ $m_{e_d}=1~{\rm MeV}, \, m_{p_d}=200~{\rm GeV} ,\, \epsilon=2.6\times 10^{-11}$]{\includegraphics[width=0.90\columnwidth]{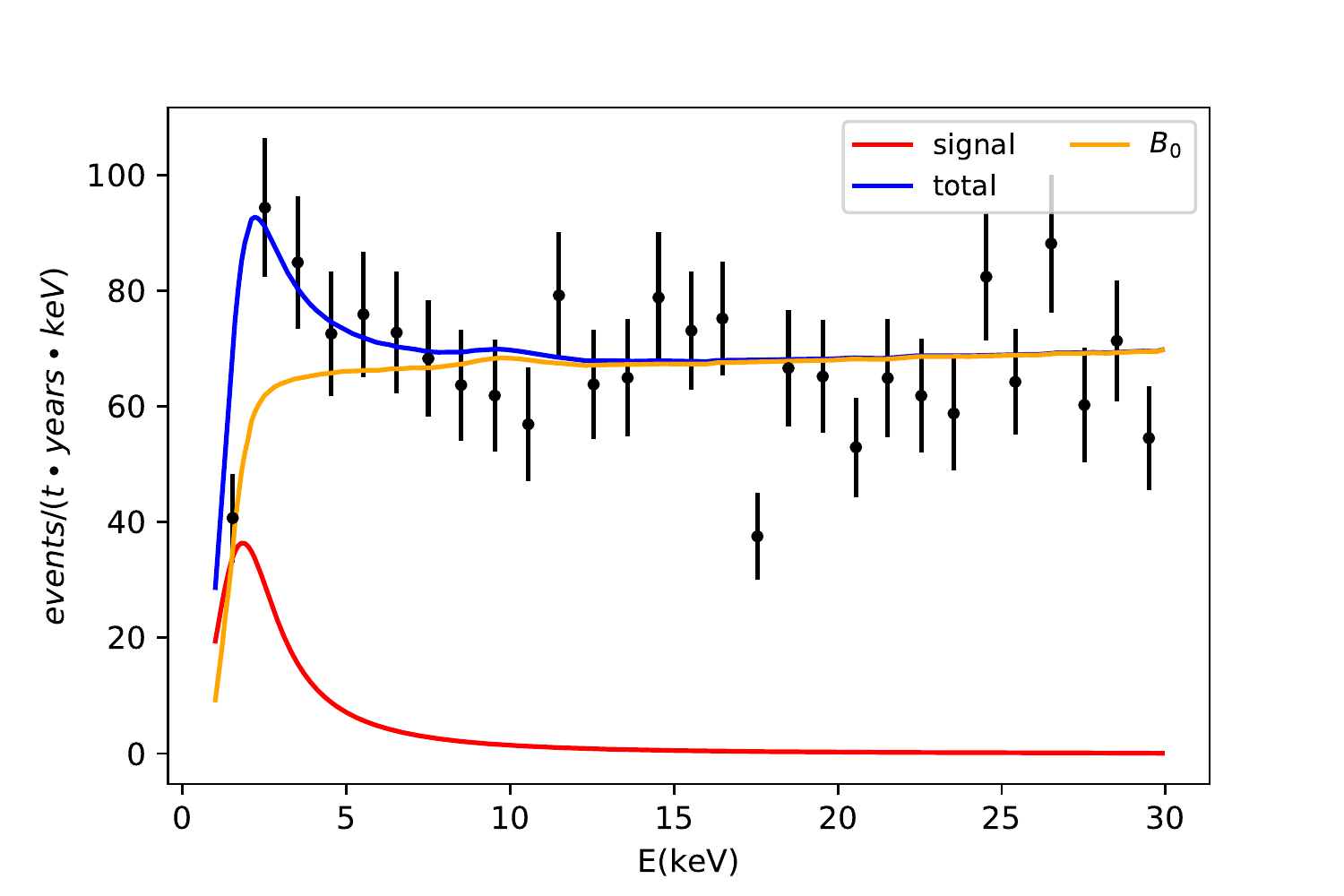}}

\subfigure[\ $m_{e_d}=1~{\rm MeV} ,\, m_{p_d}=2000~{\rm  GeV} ,\, \epsilon=1.4\times 10^{-10}$]{\includegraphics[width=0.90\columnwidth]{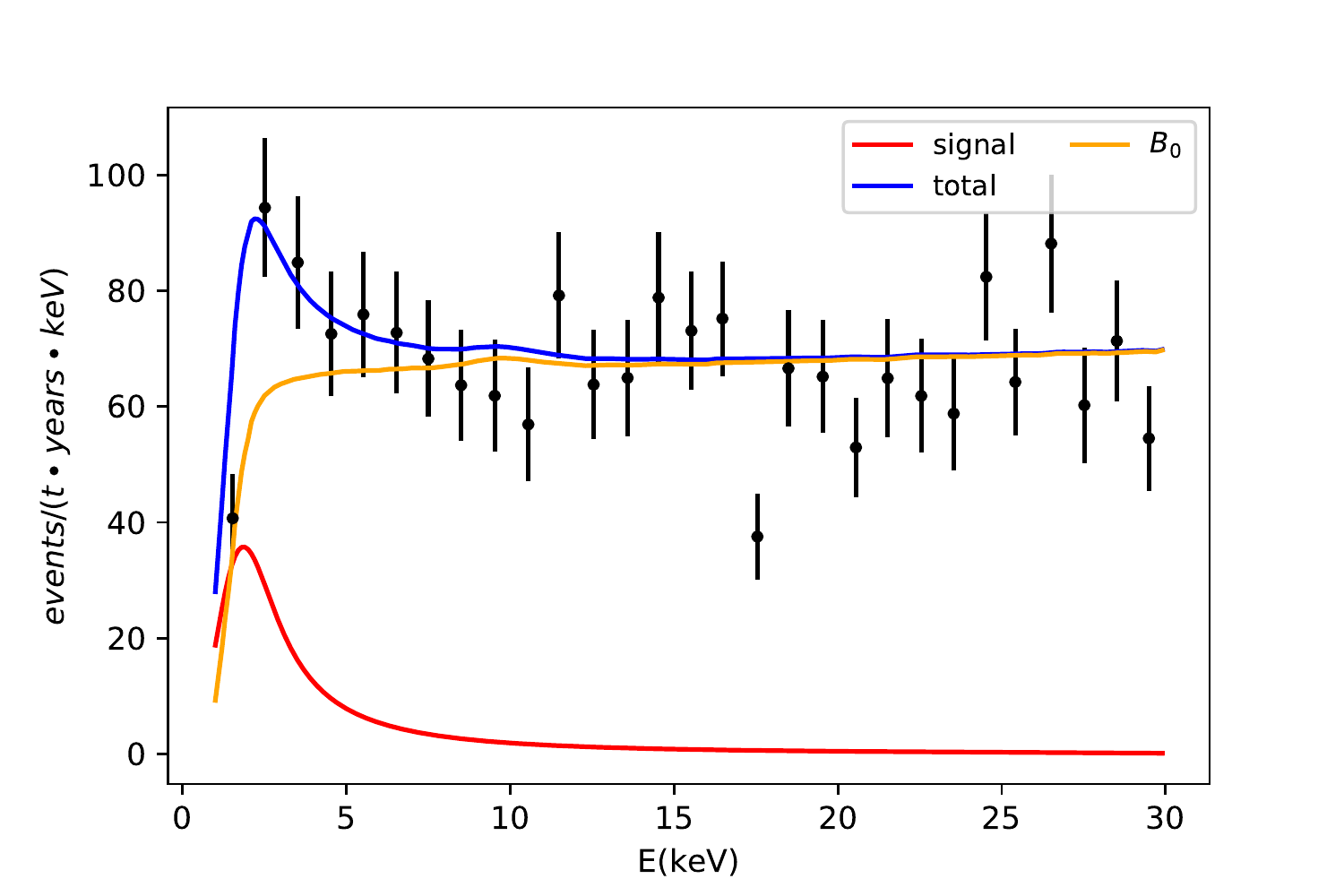}}

\caption{The plasma dark matter model prediction compared with the measured XENON1T low energy electron recoil spectrum. 
The orange line is the $B_0$ background in \cite{xenon1t}. 
The red line denotes the dark electron - electron scattering signal and the blue line presents the total events predicted by the model.}

\label{result}

\end{figure}

In the parameter region where $x \lesssim 1$, the rate has an expected $dR/dE_R \propto 1/E_R^2$ dependence,
while for $x > 1$ the rate is much steeper function of recoil energy.
Recall,
\begin{eqnarray}
x = \frac{v_{min}}{v_0} \simeq \sqrt{\frac{E_R}{m_{p_d}m_e m_{e_d}}} \frac{(m_e + m_{e_d})}{v_{rot}} .
\end{eqnarray}
Estimating $x$ numerically for $m_{e_d} \gtrsim {\rm MeV}$ gives:
\begin{eqnarray}
x \approx 0.5 \left( \frac{E_R}{{\rm keV}}\right)^{1/2} \left( \frac{20\ {\rm GeV}}{m_{p_d}}\right)^{1/2} \left( \frac{m_{e_d}}{{\rm MeV}}\right)^{1/2}
\end{eqnarray}
while for $m_{e_d} \ll {\rm MeV}$ yields:
\begin{eqnarray}
x \approx 0.2 \left( \frac{E_R}{{\rm keV}}\right)^{1/2} \left( \frac{20\ {\rm GeV}}{m_{p_d}}\right)^{1/2} \left( \frac{{\rm MeV}}{m_{e_d}}\right)^{1/2} .
\end{eqnarray}
That is, for $m_{p_d} > 20$ GeV, $x \lesssim 1$ is typically valid for the low recoil energy region of interest
$E_R \lesssim 4$ keV.

In Fig.~\ref{result} we show some examples with $m_{e_d} = 1$ MeV. The background model used here is the $B_0$ model from \cite{xenon1t}.
From the figure, we can see that this model provides a reasonable fit to the measured XENON1T electron recoil spectrum except for the
lowest energy bin. It is possible that the poor fit of the first bin may be caused by the uncertainty in modelling the rapidly falling detector efficiency 
and resolution near the threshold. Future experiments with lower energy threshold would be useful in testing
stringently the predicted $dR/dE_R \propto 1/E_R^2$ dependence.

Similar results to Fig.~\ref{result} should follow in the parameter region where $x \lesssim 1$, and the normalization suggested
by the XENON1T data constrain the coefficient of Eq.(\ref{num7}) to:
\begin{eqnarray} 
 \frac{\epsilon}{10^{-11}}  \approx 0.5 \left(\frac{{\rm MeV}}{m_{e_d}}\right)^{1/4}\left(\frac{m_{p_d}}{20 \ {\rm GeV}}\right)^{3/4} .
\end{eqnarray}
Evidently,
the kinetic mixing value needed to explain the XENON1T data lies in the approximate  range: $10^{-12} \lesssim \epsilon \lesssim 10^{-10}$. 
Such parameter space is consistent with 
cosmology and galaxy structure \cite{vagnozzi,vagnozzi2} and not ruled out by other direct experiments, such as Darkside50 \cite{Agnes}.
 
\begin{figure}[htbp]
\centering
\includegraphics[width=1\columnwidth]{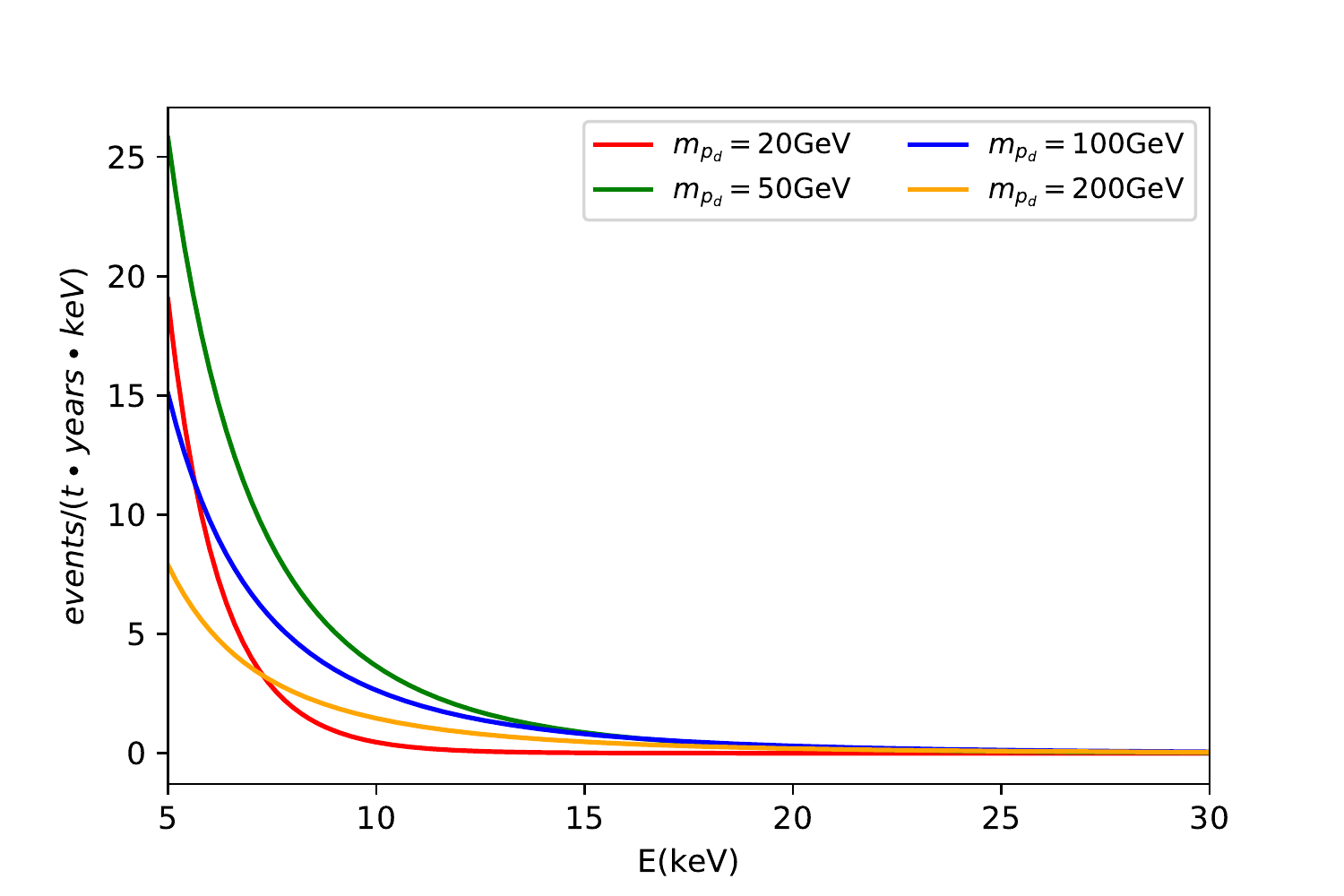}
\caption{The expected signal of dark proton - xenon nuclear scattering in an idealized detector with perfect resolution and efficiency.
The rate is proportional to $\epsilon^2$ with $\epsilon = 10^{-10}$ assumed in the figure. The different colors represent
different mass values chosen for $m_{p_d}.$ }
\label{nuclear_recoil}
\end{figure}

So far, we have focused on electron recoils due to dark electron scattering. 
The dark protons can in principle also be detected in nuclear recoil searches.
The expected rate of nuclear recoils for a xenon target is given by Eq.(\ref{drde_nuclear}).
In Fig.(\ref{nuclear_recoil}) we show the expected nuclear recoil spectrum for some illustrative parameter choices. An ideal
detector with perfect energy resolution and efficiency was assumed in this figure.

If we assume a 5 keV $E_R$ threshold (similar to that of the current XENON1T nuclear recoil analysis \cite{aprile}) then
the total expected nuclear recoil rate in the integrated energy range 5-30 keV is estimated to be
\begin{eqnarray}
R &\approx & 2.2\left(\frac{\epsilon}{3 \times 10^{-11}}\right)^2 \ {\rm t}^{-1} \ {\rm y}^{-1}  \ {\rm for}  \ m_{p_d} = 20\ {\rm GeV} \nonumber \\
R &\approx & 5.9\left(\frac{\epsilon}{3 \times 10^{-11}}\right)^2\  {\rm t}^{-1} \ {\rm y}^{-1}   \  {\rm for}  \ m_{p_d} = 50\ {\rm GeV} \nonumber \\
R &\approx & 4.0 \left(\frac{\epsilon}{3 \times 10^{-11}}\right)^2 \ {\rm t}^{-1} \ {\rm y}^{-1}\  {\rm for} \ m_{p_d} = 100\ {\rm GeV} \nonumber \\
R & \approx & 2.2\left(\frac{\epsilon}{3 \times 10^{-11}}\right)^2 \ {\rm t}^{-1} \ {\rm y}^{-1} \ {\rm for}  \ m_{p_d} = 200\ {\rm GeV} \nonumber \\
.
\end{eqnarray}
This rate is not far below current search limits, e.g. \cite{aprile}. Clearly,
dark proton interactions can be probed more rigorously in future direct detection experiments. For example,
in the XENONnT experiment \cite{xenonnt} the total nuclear recoil background in the energy range (4 keV-50 keV) is 
expected be of order 0.1 per tonne per year \cite{xenonntbg}.

\section{conclusion}

We have considered a two component dark matter model consisting of a dark proton and dark electron, each charged under an unbroken
$U(1)'$ gauge symmetry featuring also the kinetic mixing interaction. 
In this model the Milky Way halo can consist of a dark plasma.
In general, such dark matter can be captured within the Earth and strongly shield 
a detector from halo dark matter \cite{plasmadm} (mirror dark matter being a fairly
well studied case where this occurs \cite{shield1,shield2,Zu}).
However, we have explored parameter space where the physical extent of the captured dark matter  
within the Earth forms a compact sphere  (which arises for $m_{p_d} \gtrsim 20$ GeV)
and taken the idealized limiting case where any shielding effects due to captured dark matter are assumed negligible (at least for
a northern hemisphere location such as Gran Sasso).

In this kind of plasma dark matter model keV electron recoils naturally arise from the scattering of dark electrons with loosely bound atomic electrons.
We have found that these interactions
may account for the low energy electronic 
recoil spectrum observed in the XENON1T experiment. The size of the excess implicated kinetic
mixing in the approximate range: $10^{-12} \lesssim \epsilon \lesssim 10^{-10}$.
The allowed parameter space is consistent with known astrophysical and cosmological constraints (such as early Universe cosmology and supernova cooling)
and consistent also with other direct detection experiments.
It is also consistent with the kinetic mixing range suggested by small scale structure considerations \cite{vagnozzi2}.
Future experiments, including
XENONnT\cite{xenonnt}, LUX-ZEPLIN\cite{Lux}, PandaX-4T\cite{pandax}, should provide a stringent test of this explanation of the
XENON1T excess.

\vskip 0.9cm

\noindent
{\bf Acknowledgments}: YZF and FL are supported by the National Key Research
and Development Program of China (Grant No. 2016YFA0400200) and the National
Natural Science Foundation of China (Grants No. 11773075).

\end{document}